\documentclass{moriond}

\usepackage{amsmath}

\def\be{\begin{equation}}
\def\ee{\end{equation}}
\def\bea{\begin{eqnarray}}
\def\eea{\end{eqnarray}}

\newcommand{\Photo}{}

\begin{document}
\vspace*{2cm}
\title{
SELF-INTERACTING DARK MATTER AS A SOLUTION TO THE PROBLEMS IN SMALL-SCALE STRUCTURES}

\author{CAMILO GARCIA-CELY}

\address{Service de Physique Th\'eorique, Universit\'e Libre de Bruxelles, Boulevard du Triomphe, CP225,\\ 1050 Brussels, Belgium
}

\author{ XIAOYONG CHU}

\address{
Institute of High Energy Physics, Austrian Academy of Sciences, Nikolsdorfer Gasse 18,\\ 1050 Vienna, Austria
}

\maketitle\abstracts{
Dark matter self-interactions are a well-motivated
solution to the core-vs.-cusp and the too-big-to-fail problems. 
They are commonly induced by  means of  a light mediator, that is also responsible for the dark matter freeze-out in the early universe. Motivated by the fact that such scenario is excluded in its simplest realizations,  we will discuss the possibility that the relic density of a self-interacting
dark matter candidate can proceed from the freeze-out of only annihilations into
SM particles. We will argue that scalar and Majorana dark matter in the mass range of
10 to 500 MeV, coupled to a slightly heavier massive gauge boson, are the only  candidates
in agreement with multiple current experimental constraints. We will also discuss prospects of establishing or
excluding these two scenarios in future experiments.}

\section{Introduction}

The $\Lambda$CDM model is a very successful paradigm that describes a plethora of astrophysical and cosmological observations. Nevertheless, at small astrophysical scales there are a few discrepancies between its predictions and observations. Two of these discrepancies have been dubbed the core-vs.-cusp and the too-big-to-fail problems.  The former arises because N-body simulations of collisionless dark matter (DM) particles predict halos with a cusp at their center, whereas objects such as dwarf galaxies exhibit a core (See e.g. Refs.~\cite{Moore:1994yx,Flores:1994gz,deNaray:2011hy}). The second problem has to do with the fact that simulations of the Milky Way predict subhalos too massive and too dense to host the brightest observed satellites (See e.g. Refs.~\cite{BoylanKolchin:2011de,Ferrero:2011au}). 

While there is no consensus in the scientific community on the origin of these discrepancies yet, one can enumerate a couple of hypothesis. The most evident one being that the effect of baryons must be included in simulations. Possible astrophysical explanations include supernova/AGN feedback, tidal effects within halos and low star-formation rates (See Ref.~\cite{DelPopolo:2016emo} for a recent review).  One can also entertain a more  audacious alternative from the particle-physics point of view. Namely, one can postulate DM self-interactions that become relevant at small scales, without modifying the physics at large scales~\cite{Spergel:1999mh}.

The idea behind the self-interaction hypothesis is introducing a mean free path for  DM particles in halos of the order of 1 to 10 kpc. This naturally gives rise to DM distributions that are qualitatively different. Notice that in the $\Lambda$CDM paradigm, the mean free path is infinity because DM is collisionless. Given the fact that the mean free path induced by a  self-interaction cross section $\sigma$ is

\begin{equation}
\text{ Mean Free Path} \sim \left( \frac{\rho}{m_\text{DM}}\sigma\right)^{-1}\,\
\end{equation}
and taking into account the magnitude of the DM density in galactic halos, one finds  $\sigma /m_\text{DM} \sim 1 $cm$^{2}/$g at the scale of galaxies ($v \sim$ 10 - 100 km/s). Simulations show that this is indeed a plausible solution \cite{Wandelt:2000ad,Vogelsberger:2012ku,Peter:2012jh,Rocha:2012jg,Zavala:2012us,Elbert:2014bma,Kaplinghat:2015aga,Vogelsberger:2015gpr,Cyr-Racine:2015ihg}. One must keep in mind that the  order of magnitude of this cross section is much larger than the typical annihilation cross sections of DM particles produced via the freeze-out scenario. As we will see, the simplest way to achieve such difference in cross sections is by introducing a light mediator.

\section{Inducing DM-self interactions with light mediators}

\begin{figure}[t]
\label{fig:fig1}
\centering
\begin{equation}
\nonumber
\text{Light mediator } \eta
\left\{
\begin{array}{l}
\text{Long-range forces}\hspace{1.3cm}\raisebox{-0.35\height}{\includegraphics[width=0.3\textheight]{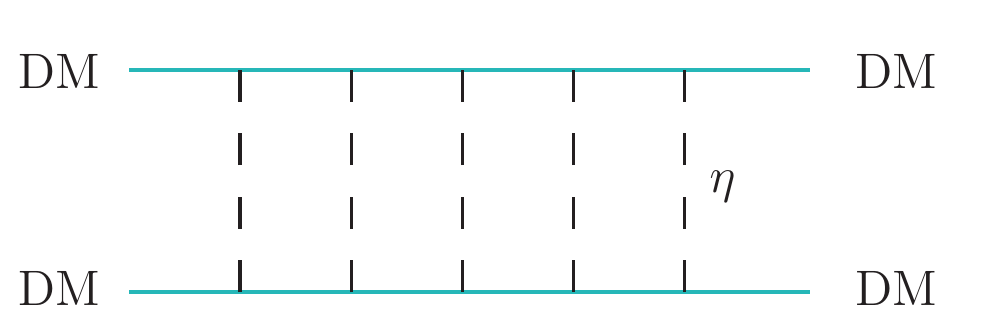}}
\\
\begin{array}{l}
\text{Freeze-out via  } \\
\sigma v_\text{ann}\sim 3\times 10^{-26} \text{cm}^3/\text{s}
\end{array}
\raisebox{-0.35\height}{\includegraphics[width=0.2\textheight]{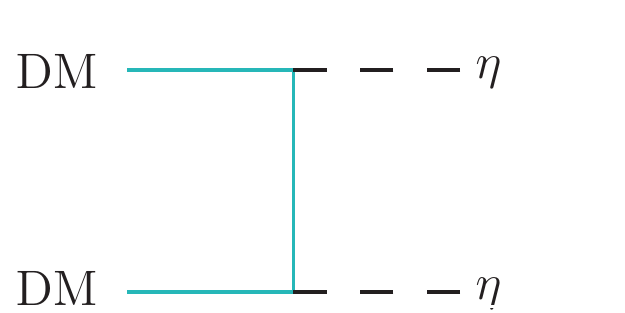}}
\end{array}
\right.
\end{equation}
\caption{Schematic representation of DM phenomenology in the presence of a light mediator.}
\end{figure}

\begin{figure}[b]
\centering
\label{fig:kinetic_equilibrium}
\includegraphics[width=0.18\textheight]{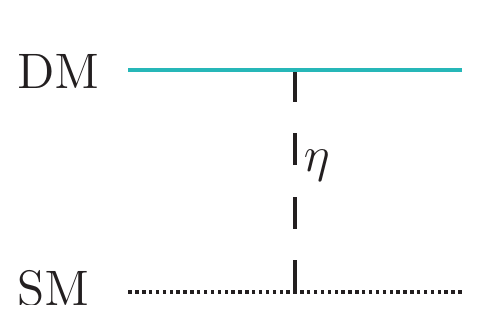}
\caption{Process establishing kinetic equilibrium between the SM and SM sectors.}
\end{figure}

One of the simplest models for self-interacting DM consists of introducing a lighter boson $\eta$  in addition to the DM particle. This particle plays two roles as summarized schematically in Fig.~1. On the one hand,  the exchange of $\eta$ bosons induces a long range force between the DM particles  in such a way that $\sigma/m_\text{DM} \sim 1 $cm$^{2}/$g. On the other hand, because of the interaction vertices $\text{DM}\,\text{DM}\, \eta$, one also naturally obtains DM annihilations into the $\eta$ particles. One can then assume that the DM is produced by means of the freeze-out of the process $\text{DM}\text{DM}\to \eta\eta$  in the early universe. Here the large difference  between the self-interaction cross section and the annihilation cross section is due to the fact that in the former case one relies on a non-perturbative effect taking place at small velocities such as the ones encountered in DM halos~\cite{Tulin:2013teo}.   
Even though this is a remarkably simple and predictive setup, in practice it is hard to achieve due to multiple constraints.

One must start by saying that the ordinary freeze-out mechanism only works if the DM and the Standard Model (SM) sectors were in thermal equilibrium at some point in the early universe. This is typically done by introducing an interaction between some SM particles and the mediator $\eta$ in such a way that the process establishing the kinetic equilibrium is the one shown in Fig.~2. Such interaction also induces the decay of the mediator alleviating possible problems with its cosmological abundance. 

Nevertheless, the presence of a light mediator coupled to the SM is severely constrained~\cite{Bernal:2015ova} because of the following reasons. First,  such mediator was present in large amounts in the early universe if it was in thermal equilibrium as required from the freeze-out mechanism. This is generally  in conflict with BBN and CMB observations if the decay of the mediator is not sufficiently fast, that is, if the coupling of the $\eta$ boson with SM particles is not sufficiently large. In addition, due to its light mass, the mediator can naturally enhance DM direct detection rates~\cite{Kaplinghat:2013yxa} if the coupling of the $\eta$ boson with SM particles is not sufficiently small. Both facts together  lead to the exclusion of many DM scenarios. Finally, due to non-perturbative effects such as the Sommerfeld enhancement, the light mediator also induces large DM annihilation signals into the mediator which subsequently decay into SM particles, affecting cosmic-ray fluxes or CMB observables~\cite{Bringmann:2016din}.   

Different avenues have been proposed in the literature to overcome the previous problems. One possibility is to consider DM production mechanisms other  than the freeze-out . This has been systematically studied in Ref.~\cite{Bernal:2015ova}. One example that fulfills all the constraints along these lines is the freeze-in mechanism. In that case, the DM or the $\eta$ particle were never in thermal equilibrium in the early universe and the DM is slowly produced from the SM bath by means of very small couplings. Notice that the abundance of the mediator is naturally much smaller, surpassing BBN and CMB bounds. This is illustrated in the left panel of Fig.~3. Interestingly, the parameters that lead to the observed abundance of DM and to its self-interaction with a strength of $1$cm${^2}$/g, can also give rise to a X-ray line with an overall flux equal to that of the tentative $3.5~$keV line~\cite{Bulbul:2014sua,Boyarsky:2014jta}.

\begin{figure}[t]
\label{fig:freezein}
\centering
\includegraphics[width=0.4\textwidth]{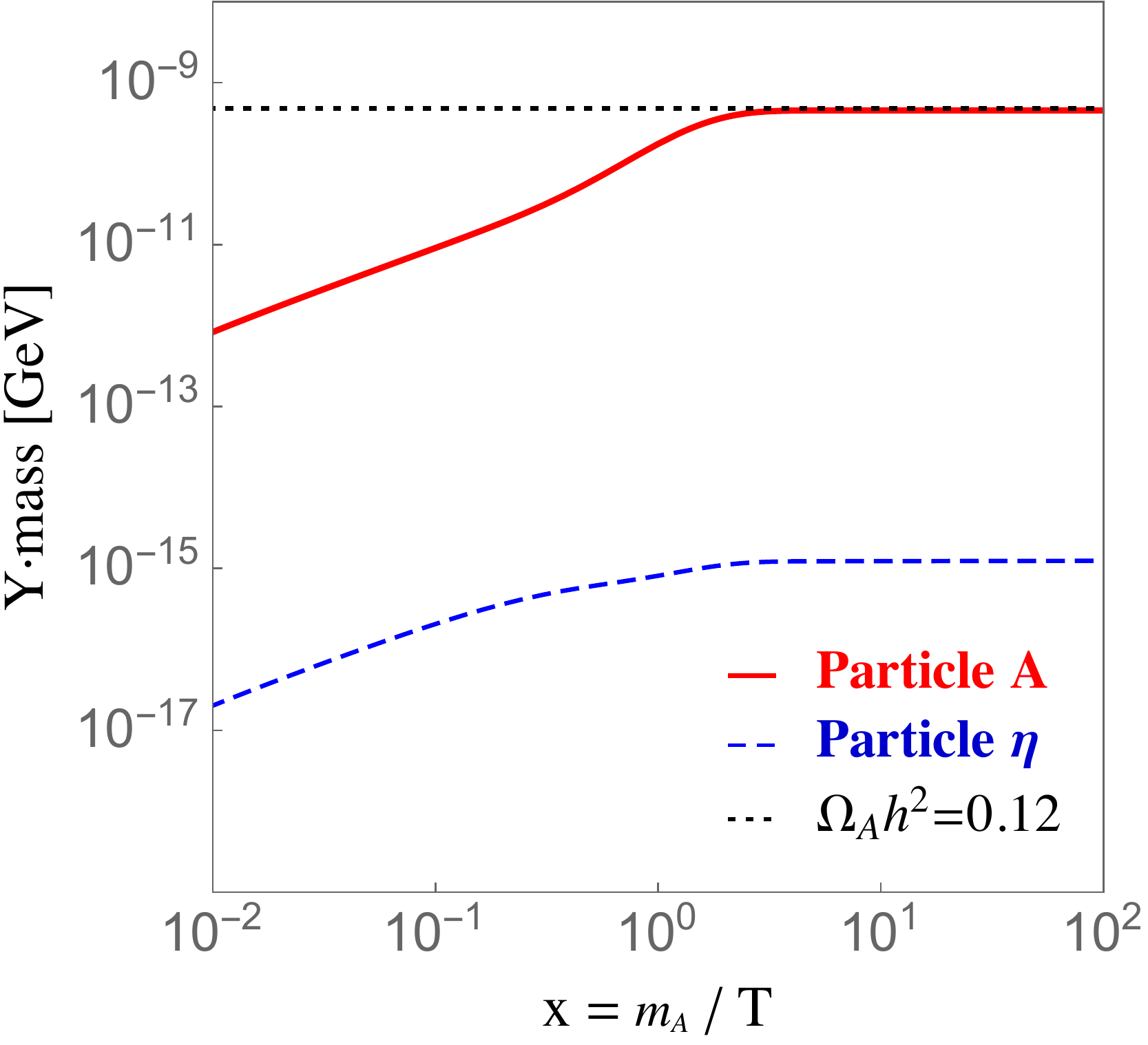}
\includegraphics[width=0.4\textwidth]{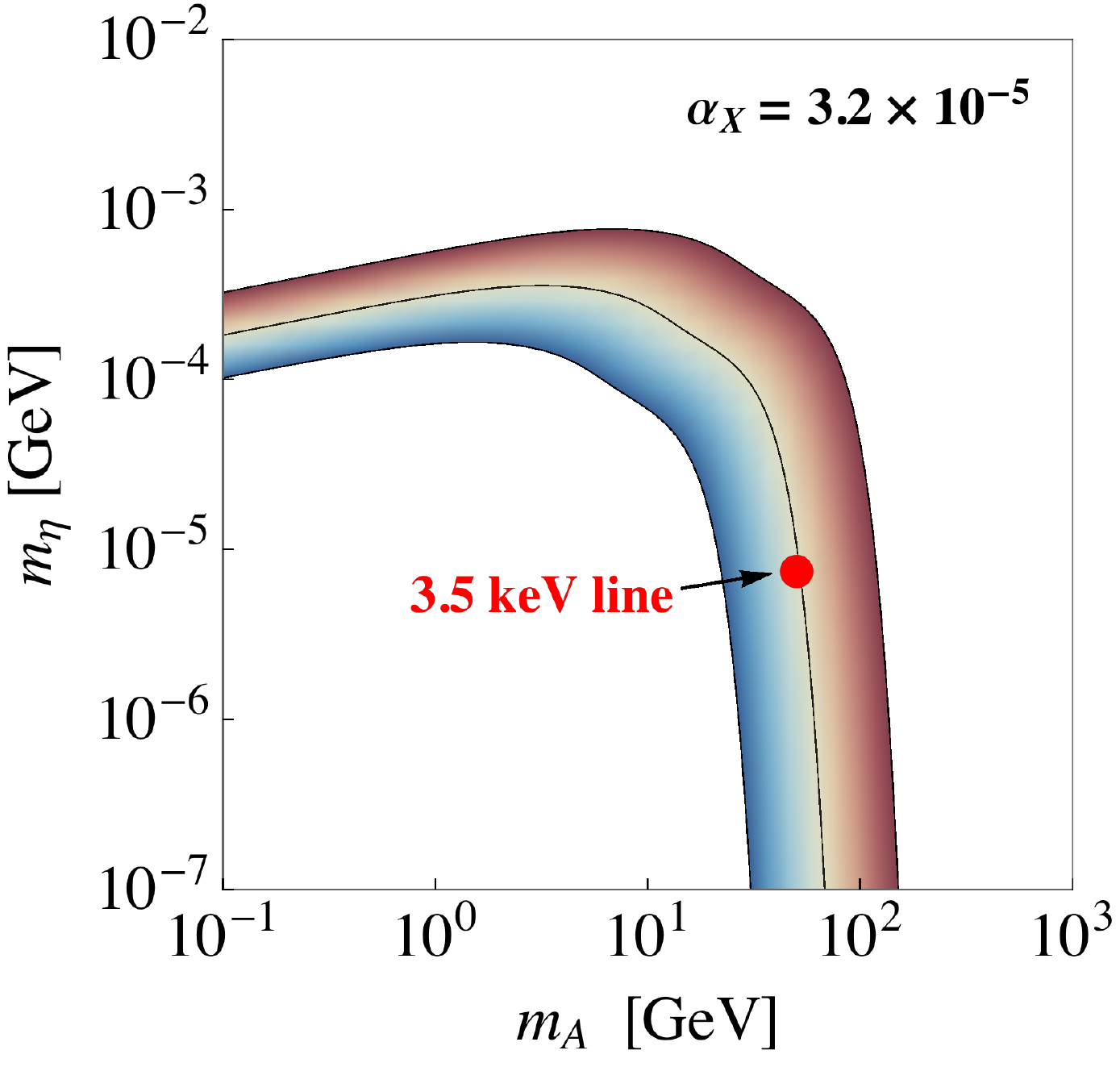}
\caption{Example of self-interacting DM produced via the freeze-in mechanism.  Here $A$ is the DM candidate. It is a vector boson with fine-structure constant $\alpha_X$. The mediator $\eta$ is a scalar boson mixing with the SM Higgs by means of an angle $\beta = 0.2\times 10^{-9}$.  \emph{Left Panel:} Abundance of DM particles as a function of SM temperature. The dotted line corresponds to the observed abundance of DM. The mediator component is naturally subdominant. \emph{Right panel:}  Parameter space giving rise to  DM self-interactions in DM halos with a velocity of 20 km/s. The contours correspond to cross sections  per DM mass of 0.1, 1 and 10 cm$^2$/g.  }
\end{figure}

\section{ Can we still consider the standard freeze-out?}

\begin{figure}[t]
\label{fig:Z'}
\centering
\includegraphics[width=0.4\textwidth]{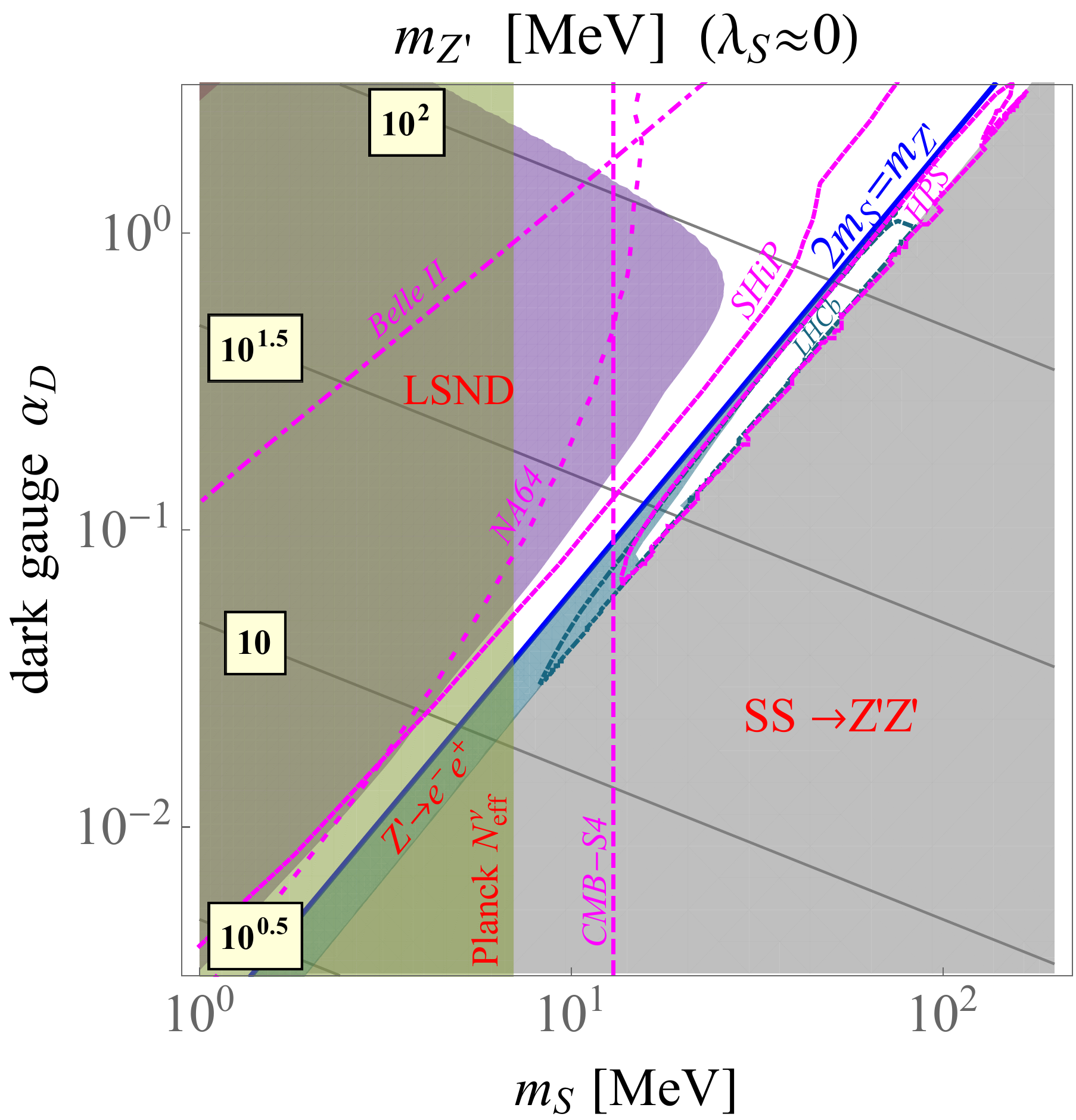}
\includegraphics[width=0.4\textwidth]{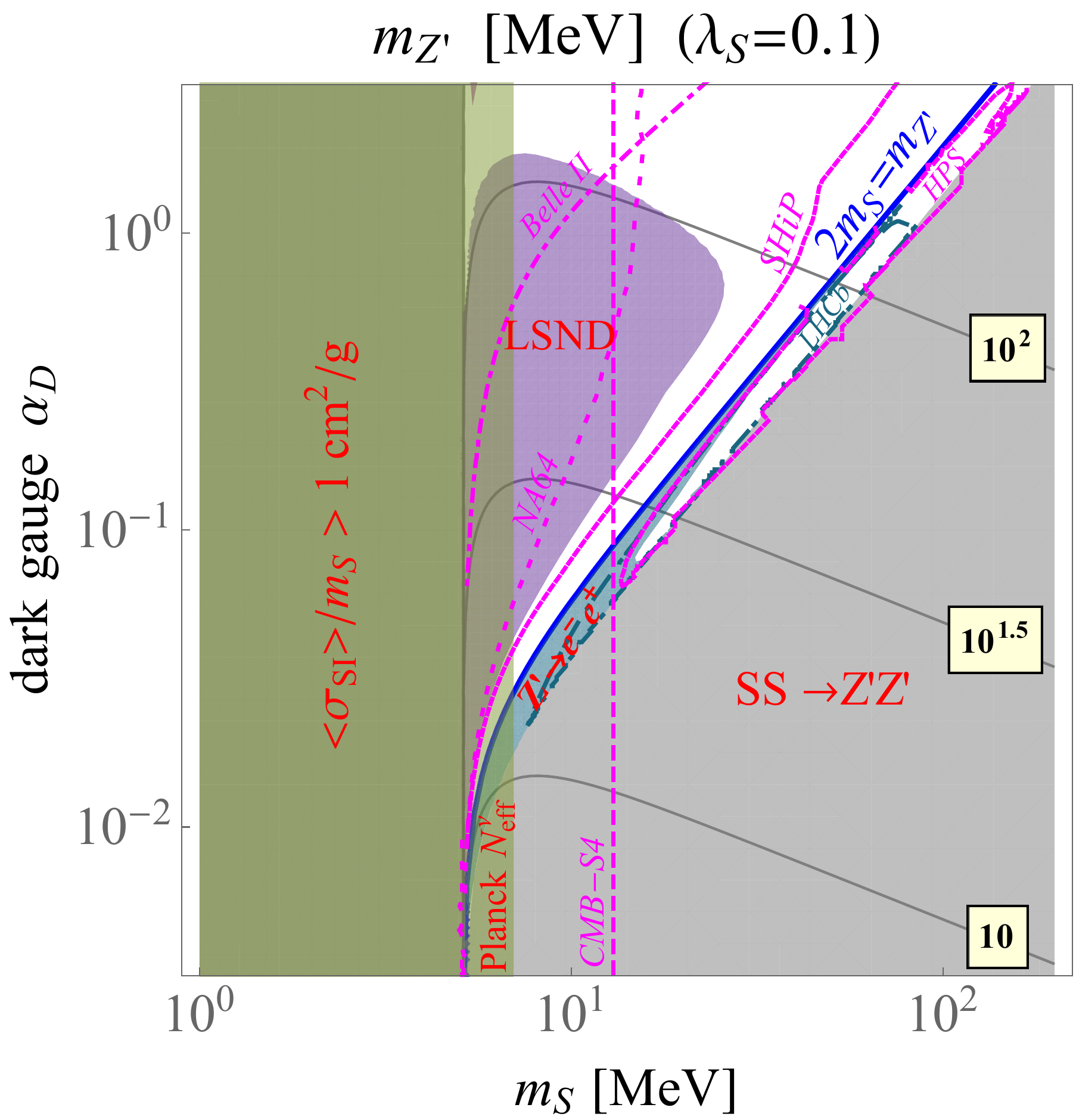}\\
\includegraphics[width=0.4\textwidth]{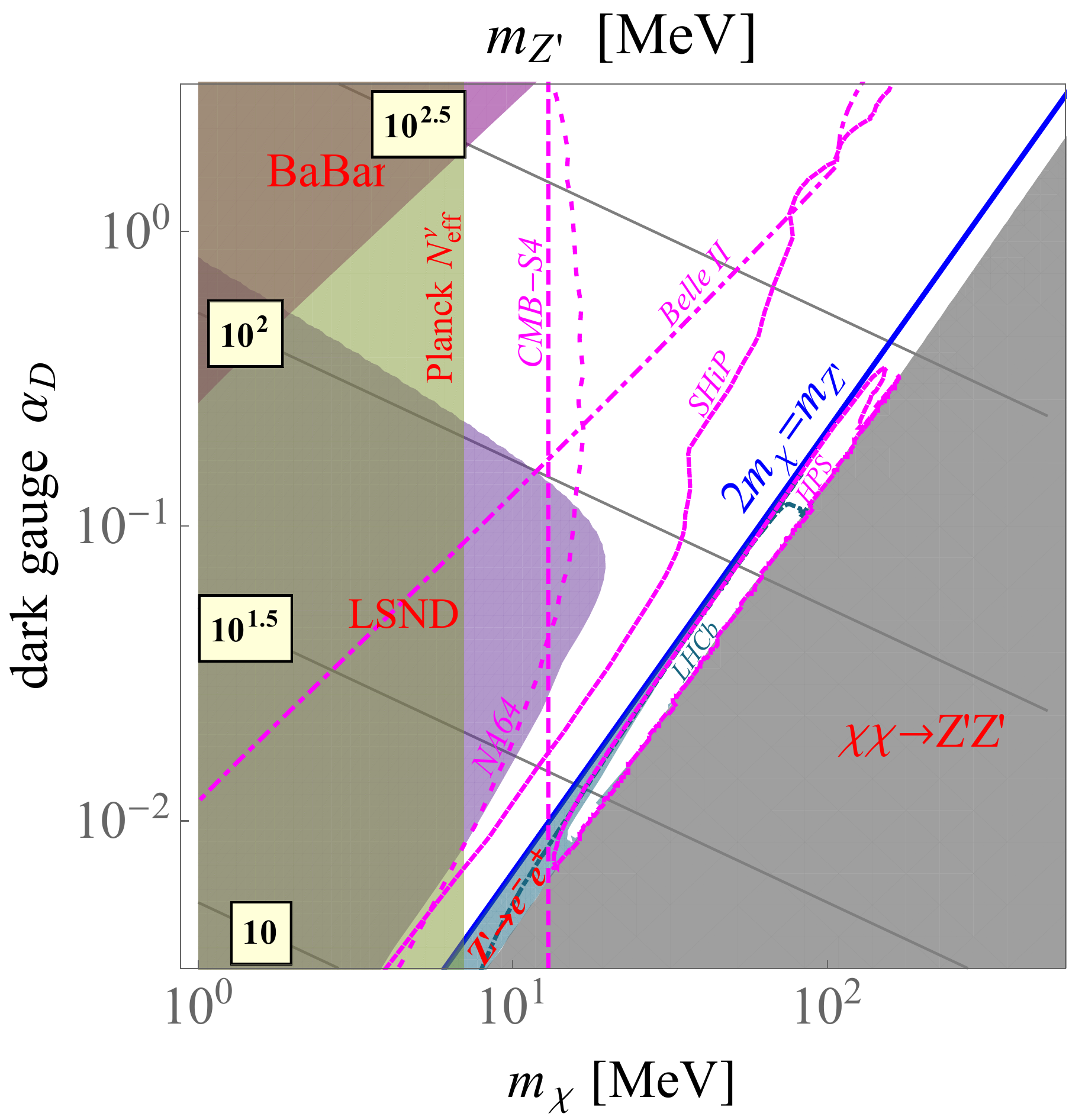}
\includegraphics[width=0.4\textwidth]{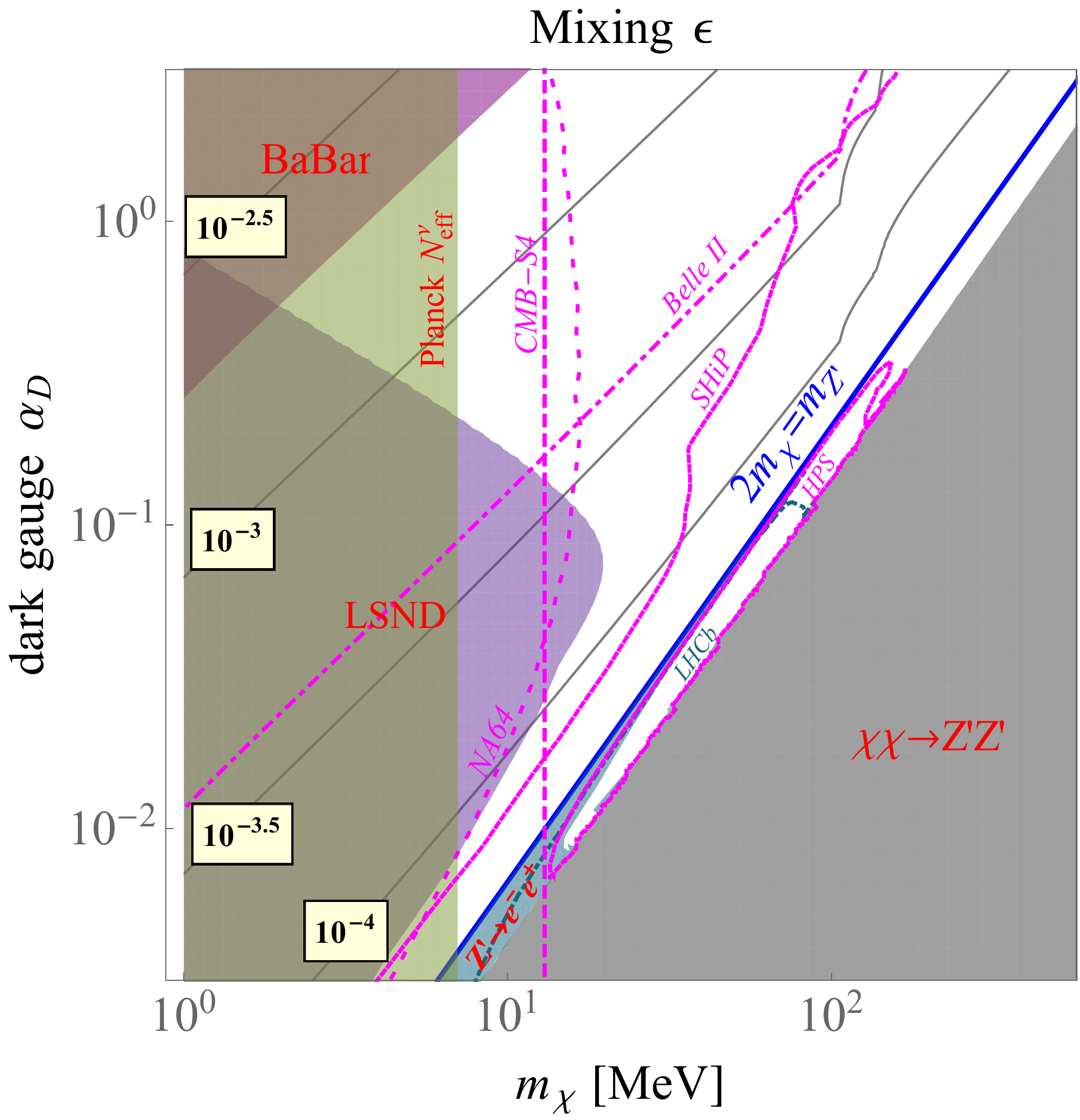}
\hspace{-1cm}
\caption{
\emph{Upper panel :  $Z'$ portal for scalar DM.}
As a function of the DM mass $m_S$ and dark coupling $\alpha_D$, the solid lines show the  $Z'$ mass satisfying the relic density and the self-interaction constraints, for two choices of the scalar self-coupling  $\lambda_S$. 
In the shaded region at \textit{right-bottom} corner, the dark freeze-out from $S S \rightarrow Z' Z'$ is too fast to account for the DM abundance.
Non-solid  (colored) lines show the expected sensitivities of future experiments. 
\emph{Lower panel:  $Z'$ portal for Majorana DM}.  As a function of the DM mass $m_\chi$ and dark coupling $\alpha_D$, the solid contour lines show the values of $Z'$ mass (\textit{left}) and kinetic mixing parameter $\epsilon$ (\textit{right}) satisfying the relic density and the self-interaction constraints.  All shaded regions are experimentally excluded in various ways. In the shaded region at the \textit{right-bottom} corner, the dark annihilation $\chi\chi\to Z'Z'$ is too fast to account for the DM abundance.   
}
\end{figure}

Part of the trouble of the freeze-out scenario discussed above  is the fact that the mediator is light. Taking this as a motivation, we now consider possible scenarios in which the mediator is heavy~\cite{Chu:2016pew}. In this case, dimensional analysis on $\sigma/{m_\text{DM}}\sim\lambda^2/m_\text{DM}^3$ directly suggests that DM must be at the MeV scale and that it must be a singlet.
We assume then that the relic density is set by means of annihilations with the  $\eta$ boson as a mediator. In addition, interactions between the mediator and the DM candidate set the self-interaction cross section.  For MeV candidates, indirect detection constraints are very strong. In fact, CMB observations rule out s-wave annihilations at the MeV scale~\cite{Slatyer:2015jla} and consequently we assume further that the annihilations are p-wave suppressed. After a systematic classification of all possible mediators, in Ref.~\cite{Chu:2016pew} we found that the mediator can only be a dark vector $Z'$ slightly heavier than the DM. 

This leads to a highly predictive and minimal scenario, in which all $Z'$ couplings to SM particles are known up to the overall multiplicative kinetic mixing parameter. Then, there are in principle four parameters: the masses of the DM and the $Z'$, the kinetic mixing $\epsilon$ and  the dark fine-structure constant $\alpha_D$. The freeze-out via $\text{DM}\,\text{DM}\to Z' \to f_\text{SM}\overline{f_\text{SM}}$ and the self-interaction hypothesis constrain two of them.

Moreover, the p-wave annihilations are realized only in two cases. These are

\begin{itemize}
\item \emph{Scalar DM coupled to a  heavier $Z'$:}
One possibility is that  DM is a scalar $S$. Annihilations into fermions are p-wave suppressed and determine the relic density via the freeze-out mechanism. Notice that there are no annihilation into photons. In addition, DM self-interactions are mediated by the exchange of the $Z'$ and/or the quartic coupling $\lambda_S$.

This is scenario is  highly testable~\cite{Izaguirre:2015pva,Alves:2015pea,Alitti:1993pn,Erler:2009jh,Essig:2013vha,Wang:2015kmm,Gninenko:2016kpg,Batell:2009di,Alekhin:2015byh,Batell:2014mga,Curtin:2014cca,Batley:2015lha,Celentano:2014wya,Ilten:2015hya,Heo:2015kra,Abazajian:2013oma,Manzotti:2015ozr} via dark photon searches and CMB observations. All of them are summarized in the upper panel of Fig.~4, where, we show the different constraints along contours of constant $Z'$ mass, for a fixed quartic coupling. See Ref.~\cite{Chu:2016pew}  for details and references to each particular experiment.  Direct detection experiments where DM scatters off electrons in semiconductors might exclude this scenario completely~\cite{Essig:2015cda}.
\item \emph{ Majorana DM coupled to a heavier $Z'$:}
The other possibility is that the DM is a Majorana fermion $\chi$. As for the previous case, annihilations into fermions are p-wave suppressed and give rise to the observed abundance of DM by means of the freeze-out mechanism. No annihilation into photons take place.  DM self-interactions are mediated by the exchange of the $Z'$ only.
As for the previous scenario, this setup is testable via dark photon searches and CMB observations.  All of them are summarized in the lower panel of Fig.~4, where  we show the different constraints along contours of constant $Z'$ mass (left) and constant kinematic mixing $\epsilon$ (right). Additional model building is required because there must be a dark scalar close in mass to the $Z'$. See Ref.~\cite{Chu:2016pew} for details.
\end{itemize}

\section{Conclusions}

Self-interacting dark matter is a well-motivated solution to the small-scale structure problems of the $\Lambda$CDM paradigm.  Multiple observations severely constrain the production of such DM candidates via the freeze-out mechanism. 
This motivates the study of other productions mechanisms such as freeze-in, as discussed in section 2. 
In this work, we have shown that freeze-out can still work if no light mediator is present and we have argued that this requires  DM to be at the MeV scale.   Furthermore, we  have seen that this is  only possible if the DM is coupled to a slightly heavier vector boson. From its simplicity and the fact that it does not require any special tuning, this scenario constitutes an 
attractive way to accommodate both DM large self-interactions and the relic density constraint. Here, the huge difference between the self-interaction and annihilation cross sections is not due to any special mechanism taking place; it is simply due to the fact that the kinetic mixing, which enters in the annihilation but not in the self-interaction, is suppressed. Moreover, this scenario offers possibilities of particle physics tests. Fig.~4 summarizes various constraints and future possibilities of testing it or ruling it out.

\section*{Acknowledgments}
We thank Nicol\'as Bernal, Thomas Hambye and Bryan Zaldivar   for collaborations on some of the work presented here. We also thank the
organizers of the Moriond EW 2017 for financial support and the opportunity to present our results.
The work of C.G.C.  is supported by the FNRS, by the IISN and by the 
Belgian Federal Science Policy through the Interuniversity Attraction Pole P7/37.


\section*{References}

\begin{thebibliography}{10}

\bibitem{Moore:1994yx}
B.~Moore, ``{\em {Evidence against dissipationless dark matter from
  observations of galaxy haloes}},''
\href{http://dx.doi.org/10.1038/370629a0}{Nature {\normalfont \bfseries 370}
  (1994)  629}.

\bibitem{Flores:1994gz}
R.~A. Flores and J.~R. Primack, ``{\em {Observational and theoretical
  constraints on singular dark matter halos}},''
  \href{http://dx.doi.org/10.1086/187350}{Astrophys. J. {\normalfont \bfseries
  427} (1994)  L1--4},
\href{http://arxiv.org/abs/astro-ph/9402004}{{\normalfont \ttfamily
  arXiv:astro-ph/9402004}}.

\bibitem{deNaray:2011hy}
R.~Kuzio~de Naray and K.~Spekkens, ``{\em {Do Baryons Alter the Halos of Low
  Surface Brightness Galaxies?}},''
  \href{http://dx.doi.org/10.1088/2041-8205/741/2/L29}{Astrophys. J.
  {\normalfont \bfseries 741} (2011)  L29},
\href{http://arxiv.org/abs/1109.1288}{{\normalfont \ttfamily arXiv:1109.1288}}.

\bibitem{BoylanKolchin:2011de}
M.~Boylan-Kolchin, J.~S. Bullock, and M.~Kaplinghat, ``{\em {Too big to fail?
  The puzzling darkness of massive Milky Way subhaloes}},''
  \href{http://dx.doi.org/10.1111/j.1745-3933.2011.01074.x}{Mon. Not. Roy.
  Astron. Soc. {\normalfont \bfseries 415} (2011)  L40},
\href{http://arxiv.org/abs/1103.0007}{{\normalfont \ttfamily arXiv:1103.0007}}.

\bibitem{Ferrero:2011au}
I.~Ferrero, M.~G. Abadi, J.~F. Navarro, L.~V. Sales, and S.~Gurovich, ``{\em
  {The dark matter halos of dwarf galaxies: a challenge for the LCDM
  paradigm?}},'' \href{http://dx.doi.org/10.1111/j.1365-2966.2012.21623.x}{Mon.
  Not. Roy. Astron. Soc. {\normalfont \bfseries 425} (2012)  2817--2823},
\href{http://arxiv.org/abs/1111.6609}{{\normalfont \ttfamily arXiv:1111.6609}}.

\bibitem{DelPopolo:2016emo}
A.~Del~Popolo and M.~Le~Delliou, ``{\em {Small scale problems of the
  $\Lambda$CDM model: a short review}},''
  \href{http://dx.doi.org/10.3390/galaxies5010017}{Galaxies {\normalfont
  \bfseries 5} (2017) no.~1, 17},
\href{http://arxiv.org/abs/1606.07790}{{\normalfont \ttfamily
  arXiv:1606.07790}}.

\bibitem{Spergel:1999mh}
D.~N. Spergel and P.~J. Steinhardt, ``{\em {Observational evidence for
  selfinteracting cold dark matter}},''
  \href{http://dx.doi.org/10.1103/PhysRevLett.84.3760}{Phys. Rev. Lett.
  {\normalfont \bfseries 84} (2000)  3760--3763},
\href{http://arxiv.org/abs/astro-ph/9909386}{{\normalfont \ttfamily
  arXiv:astro-ph/9909386}}.

\bibitem{Wandelt:2000ad}
B.~D. Wandelt, R.~Dave, G.~R. Farrar, P.~C. McGuire, D.~N. Spergel, and P.~J.
  Steinhardt, ``{\em {Selfinteracting dark matter}},'' in {\em {Sources and
  detection of dark matter and dark energy in the universe. Proceedings, 4th
  International Symposium, DM 2000, Marina del Rey, USA, February 23-25,
  2000}}, pp.~263--274.
\newblock 2000.
\newblock \href{http://arxiv.org/abs/astro-ph/0006344}{{\normalfont \ttfamily
  arXiv:astro-ph/0006344}}.
\newblock
\url{http://www.slac.stanford.edu/spires/find/books/www?cl=QB461:I57:2000}.
\newblock

\bibitem{Vogelsberger:2012ku}
M.~Vogelsberger, J.~Zavala, and A.~Loeb, ``{\em {Subhaloes in Self-Interacting
  Galactic Dark Matter Haloes}},''
  \href{http://dx.doi.org/10.1111/j.1365-2966.2012.21182.x}{Mon. Not. Roy.
  Astron. Soc. {\normalfont \bfseries 423} (2012)  3740},
\href{http://arxiv.org/abs/1201.5892}{{\normalfont \ttfamily arXiv:1201.5892}}.

\bibitem{Peter:2012jh}
A.~H.~G. Peter, M.~Rocha, J.~S. Bullock, and M.~Kaplinghat, ``{\em
  {Cosmological Simulations with Self-Interacting Dark Matter II: Halo Shapes
  vs. Observations}},'' \href{http://dx.doi.org/10.1093/mnras/sts535}{Mon. Not.
  Roy. Astron. Soc. {\normalfont \bfseries 430} (2013)  105},
\href{http://arxiv.org/abs/1208.3026}{{\normalfont \ttfamily arXiv:1208.3026}}.

\bibitem{Rocha:2012jg}
M.~Rocha, A.~H.~G. Peter, J.~S. Bullock, M.~Kaplinghat, S.~Garrison-Kimmel,
  J.~Onorbe, and L.~A. Moustakas, ``{\em {Cosmological Simulations with
  Self-Interacting Dark Matter I: Constant Density Cores and Substructure}},''
  \href{http://dx.doi.org/10.1093/mnras/sts514}{Mon. Not. Roy. Astron. Soc.
  {\normalfont \bfseries 430} (2013)  81--104},
\href{http://arxiv.org/abs/1208.3025}{{\normalfont \ttfamily arXiv:1208.3025}}.

\bibitem{Zavala:2012us}
J.~Zavala, M.~Vogelsberger, and M.~G. Walker, ``{\em {Constraining
  Self-Interacting Dark Matter with the Milky Way's dwarf spheroidals}},''
  \href{http://dx.doi.org/10.1093/mnrasl/sls053}{Monthly Notices of the Royal
  Astronomical Society: Letters {\normalfont \bfseries 431} (2013)  L20--L24},
\href{http://arxiv.org/abs/1211.6426}{{\normalfont \ttfamily arXiv:1211.6426}}.

\bibitem{Elbert:2014bma}
O.~D. Elbert, J.~S. Bullock, S.~Garrison-Kimmel, M.~Rocha, J.~Oñorbe, and
  A.~H.~G. Peter, ``{\em {Core formation in dwarf haloes with self-interacting
  dark matter: no fine-tuning necessary}},''
  \href{http://dx.doi.org/10.1093/mnras/stv1470}{Mon. Not. Roy. Astron. Soc.
  {\normalfont \bfseries 453} (2015) no.~1, 29--37},
\href{http://arxiv.org/abs/1412.1477}{{\normalfont \ttfamily arXiv:1412.1477}}.

\bibitem{Kaplinghat:2015aga}
M.~Kaplinghat, S.~Tulin, and H.-B. Yu, ``{\em {Dark Matter Halos as Particle
  Colliders: Unified Solution to Small-Scale Structure Puzzles from Dwarfs to
  Clusters}},'' \href{http://dx.doi.org/10.1103/PhysRevLett.116.041302}{Phys.
  Rev. Lett. {\normalfont \bfseries 116} (2016) no.~4, 041302},
\href{http://arxiv.org/abs/1508.03339}{{\normalfont \ttfamily
  arXiv:1508.03339}}.

\bibitem{Vogelsberger:2015gpr}
M.~Vogelsberger, J.~Zavala, F.-Y. Cyr-Racine, C.~Pfrommer, T.~Bringmann, and
  K.~Sigurdson, ``{\em {ETHOS – an effective theory of structure formation:
  dark matter physics as a possible explanation of the small-scale CDM
  problems}},'' \href{http://dx.doi.org/10.1093/mnras/stw1076}{Mon. Not. Roy.
  Astron. Soc. {\normalfont \bfseries 460} (2016) no.~2, 1399--1416},
\href{http://arxiv.org/abs/1512.05349}{{\normalfont \ttfamily
  arXiv:1512.05349}}.

\bibitem{Cyr-Racine:2015ihg}
F.-Y. Cyr-Racine, K.~Sigurdson, J.~Zavala, T.~Bringmann, M.~Vogelsberger, and
  C.~Pfrommer, ``{\em {ETHOS—an effective theory of structure formation: From
  dark particle physics to the matter distribution of the Universe}},''
  \href{http://dx.doi.org/10.1103/PhysRevD.93.123527}{Phys. Rev. {\normalfont
  \bfseries D93} (2016) no.~12, 123527},
\href{http://arxiv.org/abs/1512.05344}{{\normalfont \ttfamily
  arXiv:1512.05344}}.

\bibitem{Bernal:2015ova}
N.~Bernal, X.~Chu, C.~Garcia-Cely, T.~Hambye, and B.~Zaldivar, ``{\em
  {Production Regimes for Self-Interacting Dark Matter}},''
  \href{http://dx.doi.org/10.1088/1475-7516/2016/03/018}{JCAP {\normalfont
  \bfseries 1603} (2016) no.~03, 018},
\href{http://arxiv.org/abs/1510.08063}{{\normalfont \ttfamily
  arXiv:1510.08063}}.

\bibitem{Kaplinghat:2013yxa}
M.~Kaplinghat, S.~Tulin, and H.-B. Yu, ``{\em {Direct Detection Portals for
  Self-interacting Dark Matter}},''
  \href{http://dx.doi.org/10.1103/PhysRevD.89.035009}{Phys. Rev. {\normalfont
  \bfseries D89} (2014) no.~3, 035009},
\href{http://arxiv.org/abs/1310.7945}{{\normalfont \ttfamily arXiv:1310.7945}}.

\bibitem{Bringmann:2016din}
T.~Bringmann, F.~Kahlhoefer, K.~Schmidt-Hoberg, and P.~Walia, ``{\em {Strong
  constraints on self-interacting dark matter with light mediators}},''
  \href{http://dx.doi.org/10.1103/PhysRevLett.118.141802}{Phys. Rev. Lett.
  {\normalfont \bfseries 118} (2017) no.~14, 141802},
\href{http://arxiv.org/abs/1612.00845}{{\normalfont \ttfamily
  arXiv:1612.00845}}.

\bibitem{Bulbul:2014sua}
E.~Bulbul, M.~Markevitch, A.~Foster, R.~K. Smith, M.~Loewenstein, and S.~W.
  Randall, ``{\em {Detection of An Unidentified Emission Line in the Stacked
  X-ray spectrum of Galaxy Clusters}},''
  \href{http://dx.doi.org/10.1088/0004-637X/789/1/13}{Astrophys. J.
  {\normalfont \bfseries 789} (2014)  13},
\href{http://arxiv.org/abs/1402.2301}{{\normalfont \ttfamily arXiv:1402.2301}}.

\bibitem{Boyarsky:2014jta}
A.~Boyarsky, O.~Ruchayskiy, D.~Iakubovskyi, and J.~Franse, ``{\em {Unidentified
  Line in X-Ray Spectra of the Andromeda Galaxy and Perseus Galaxy Cluster}},''
  \href{http://dx.doi.org/10.1103/PhysRevLett.113.251301}{Phys. Rev. Lett.
  {\normalfont \bfseries 113} (2014)  251301},
\href{http://arxiv.org/abs/1402.4119}{{\normalfont \ttfamily arXiv:1402.4119}}.

\bibitem{Chu:2016pew}
X.~Chu, C.~Garcia-Cely, and T.~Hambye, ``{\em {Can the relic density of
  self-interacting dark matter be due to annihilations into Standard Model
  particles?}},'' \href{http://dx.doi.org/10.1007/JHEP11(2016)048}{JHEP
  {\normalfont \bfseries 11} (2016)  048},
\href{http://arxiv.org/abs/1609.00399}{{\normalfont \ttfamily
  arXiv:1609.00399}}.

\bibitem{Slatyer:2015jla}
T.~R. Slatyer, ``{\em {Indirect dark matter signatures in the cosmic dark ages.
  I. Generalizing the bound on s-wave dark matter annihilation from Planck
  results}},'' \href{http://dx.doi.org/10.1103/PhysRevD.93.023527}{Phys. Rev.
  {\normalfont \bfseries D93} (2016) no.~2, 023527},
\href{http://arxiv.org/abs/1506.03811}{{\normalfont \ttfamily
  arXiv:1506.03811}}.





\bibitem{Tulin:2013teo}
S.~Tulin, H.-B. Yu, and K.~M. Zurek, ``{\em {Beyond Collisionless Dark Matter:
  Particle Physics Dynamics for Dark Matter Halo Structure}},''
  \href{http://dx.doi.org/10.1103/PhysRevD.87.115007}{Phys.Rev. {\normalfont
  \bfseries D87} (2013) no.~11, 115007},
\href{http://arxiv.org/abs/1302.3898}{{\normalfont \ttfamily arXiv:1302.3898}}.



\bibitem{Izaguirre:2015pva}
E.~Izaguirre, G.~Krnjaic, and M.~Pospelov, ``{\em {MeV-Scale Dark Matter Deep
  Underground}},'' \href{http://dx.doi.org/10.1103/PhysRevD.92.095014}{Phys.
  Rev. {\normalfont \bfseries D92} (2015) no.~9, 095014},
\href{http://arxiv.org/abs/1507.02681}{{\normalfont \ttfamily
  arXiv:1507.02681}}.

\bibitem{Alves:2015pea}
A.~Alves, A.~Berlin, S.~Profumo, and F.~S. Queiroz, ``{\em {Dark Matter
  Complementarity and the Z$^\prime$ Portal}},''
  \href{http://dx.doi.org/10.1103/PhysRevD.92.083004}{Phys. Rev. {\normalfont
  \bfseries D92} (2015) no.~8, 083004},
\href{http://arxiv.org/abs/1501.03490}{{\normalfont \ttfamily
  arXiv:1501.03490}}.

\bibitem{Alitti:1993pn}
{\normalfont \bfseries UA2}, J.~Alitti {\em et al.}, ``{\em {A Search for new
  intermediate vector mesons and excited quarks decaying to two jets at the
  CERN $\bar{p} p$ collider}},''
\href{http://dx.doi.org/10.1016/0550-3213(93)90395-6}{Nucl. Phys. {\normalfont
  \bfseries B400} (1993)  3--24}.

\bibitem{Erler:2009jh}
J.~Erler, P.~Langacker, S.~Munir, and E.~Rojas, ``{\em {Improved Constraints on
  Z-prime Bosons from Electroweak Precision Data}},''
  \href{http://dx.doi.org/10.1088/1126-6708/2009/08/017}{JHEP {\normalfont
  \bfseries 08} (2009)  017},
\href{http://arxiv.org/abs/0906.2435}{{\normalfont \ttfamily arXiv:0906.2435}}.


\bibitem{Essig:2013vha}
R.~Essig, J.~Mardon, M.~Papucci, T.~Volansky, and Y.-M. Zhong, ``{\em
  {Constraining Light Dark Matter with Low-Energy $e^+e^-$ Colliders}},''
  \href{http://dx.doi.org/10.1007/JHEP11(2013)167}{JHEP {\normalfont \bfseries
  11} (2013)  167},
\href{http://arxiv.org/abs/1309.5084}{{\normalfont \ttfamily arXiv:1309.5084}}.

\bibitem{Wang:2015kmm}
{\normalfont \bfseries Belle-II}, B.~Wang, ``{\em {The Belle II Experiment and
  SuperKEKB Upgrade}},'' in {\em {10th International Workshop on e+e-
  collisions from Phi to Psi (PHIPSI15) Hefei, Anhui, China, September 23-26,
  2015}}.
\newblock 2015.
\newblock \href{http://arxiv.org/abs/1511.09434}{{\normalfont \ttfamily
  arXiv:1511.09434}}.
\newblock
\url{https://inspirehep.net/record/1407151/files/arXiv:1511.09434.pdf}.
\newblock


\bibitem{Gninenko:2016kpg}
S.~N. Gninenko, N.~V. Krasnikov, M.~M. Kirsanov, and D.~V. Kirpichnikov, ``{\em
  {Missing energy signature from invisible decays of dark photons at the CERN
  SPS}},''
\href{http://arxiv.org/abs/1604.08432}{{\normalfont \ttfamily
  arXiv:1604.08432}}.

\bibitem{Batell:2009di}
B.~Batell, M.~Pospelov, and A.~Ritz, ``{\em {Exploring Portals to a Hidden
  Sector Through Fixed Targets}},''
  \href{http://dx.doi.org/10.1103/PhysRevD.80.095024}{Phys. Rev. {\normalfont
  \bfseries D80} (2009)  095024},
\href{http://arxiv.org/abs/0906.5614}{{\normalfont \ttfamily arXiv:0906.5614}}.

\bibitem{Alekhin:2015byh}
S.~Alekhin {\em et al.}, ``{\em {A facility to Search for Hidden Particles at
  the CERN SPS: the SHiP physics case}},''
\href{http://arxiv.org/abs/1504.04855}{{\normalfont \ttfamily
  arXiv:1504.04855}}.

\bibitem{Batell:2014mga}
B.~Batell, R.~Essig, and Z.~Surujon, ``{\em {Strong Constraints on Sub-GeV Dark
  Sectors from SLAC Beam Dump E137}},''
  \href{http://dx.doi.org/10.1103/PhysRevLett.113.171802}{Phys. Rev. Lett.
  {\normalfont \bfseries 113} (2014) no.~17, 171802},
\href{http://arxiv.org/abs/1406.2698}{{\normalfont \ttfamily arXiv:1406.2698}}.




\bibitem{Curtin:2014cca}
D.~Curtin, R.~Essig, S.~Gori, and J.~Shelton, ``{\em {Illuminating Dark Photons
  with High-Energy Colliders}},''
  \href{http://dx.doi.org/10.1007/JHEP02(2015)157}{JHEP {\normalfont \bfseries
  02} (2015)  157},
\href{http://arxiv.org/abs/1412.0018}{{\normalfont \ttfamily arXiv:1412.0018}}.



\bibitem{Batley:2015lha}
{\normalfont \bfseries NA48/2}, J.~R. Batley {\em et al.}, ``{\em {Search for
  the dark photon in $\pi^0$ decays}},''
  \href{http://dx.doi.org/10.1016/j.physletb.2015.04.068}{Phys. Lett.
  {\normalfont \bfseries B746} (2015)  178--185},
\href{http://arxiv.org/abs/1504.00607}{{\normalfont \ttfamily
  arXiv:1504.00607}}.


\bibitem{Celentano:2014wya}
{\normalfont \bfseries HPS}, A.~Celentano, ``{\em {The Heavy Photon Search
  experiment at Jefferson Laboratory}},''
  \href{http://dx.doi.org/10.1088/1742-6596/556/1/012064}{J. Phys. Conf. Ser.
  {\normalfont \bfseries 556} (2014) no.~1, 012064},
\href{http://arxiv.org/abs/1505.02025}{{\normalfont \ttfamily
  arXiv:1505.02025}}.




\bibitem{Ilten:2015hya}
P.~Ilten, J.~Thaler, M.~Williams, and W.~Xue, ``{\em {Dark photons from charm
  mesons at LHCb}},'' \href{http://dx.doi.org/10.1103/PhysRevD.92.115017}{Phys.
  Rev. {\normalfont \bfseries D92} (2015) no.~11, 115017},
\href{http://arxiv.org/abs/1509.06765}{{\normalfont \ttfamily
  arXiv:1509.06765}}.


\bibitem{Heo:2015kra}
J.~H. Heo and C.~S. Kim, ``{\em {Light Dark Matter and Dark Radiation}},''
  \href{http://dx.doi.org/10.3938/jkps.68.715}{J. Korean Phys. Soc.
  {\normalfont \bfseries 68} (2016) no.~5, 715--721},
\href{http://arxiv.org/abs/1504.00773}{{\normalfont \ttfamily
  arXiv:1504.00773}}.

\bibitem{Abazajian:2013oma}
{\normalfont \bfseries Topical Conveners: K.N. Abazajian, J.E. Carlstrom, A.T.
  Lee}, K.~N. Abazajian {\em et al.}, ``{\em {Neutrino Physics from the Cosmic
  Microwave Background and Large Scale Structure}},''
  \href{http://dx.doi.org/10.1016/j.astropartphys.2014.05.014}{Astropart. Phys.
  {\normalfont \bfseries 63} (2015)  66--80},
\href{http://arxiv.org/abs/1309.5383}{{\normalfont \ttfamily arXiv:1309.5383}}.

\bibitem{Manzotti:2015ozr}
A.~Manzotti, S.~Dodelson, and Y.~Park, ``{\em {External priors for the next
  generation of CMB experiments}},''
  \href{http://dx.doi.org/10.1103/PhysRevD.93.063009}{Phys. Rev. {\normalfont
  \bfseries D93} (2016) no.~6, 063009},
\href{http://arxiv.org/abs/1512.02654}{{\normalfont \ttfamily
  arXiv:1512.02654}}.

\bibitem{Fradette:2014sza}
A.~Fradette, M.~Pospelov, J.~Pradler, and A.~Ritz, ``{\em {Cosmological
  Constraints on Very Dark Photons}},''
  \href{http://dx.doi.org/10.1103/PhysRevD.90.035022}{Phys. Rev. {\normalfont
  \bfseries D90} (2014) no.~3, 035022},
\href{http://arxiv.org/abs/1407.0993}{{\normalfont \ttfamily arXiv:1407.0993}}.

\bibitem{Essig:2015cda}
R.~Essig, M.~Fernandez-Serra, J.~Mardon, A.~Soto, T.~Volansky, and T.-T. Yu,
  ``{\em {Direct Detection of sub-GeV Dark Matter with Semiconductor
  Targets}},'' \href{http://dx.doi.org/10.1007/JHEP05(2016)046}{JHEP
  {\normalfont \bfseries 05} (2016)  046},
\href{http://arxiv.org/abs/1509.01598}{{\normalfont \ttfamily
  arXiv:1509.01598}}.



\end{thebibliography}
\end{document}